# Physics-driven Deep Learning Inversion for Direct Current Resistivity Survey Data

Bin Liu[1,2], Yonghao Pang[1,3], Peng Jiang[1,3], Zhengyu Liu[1,2], Benchao Liu[1,2], Yongheng Zhang[1,3], Yumei Cai[1,3], Jiawen Liu[1,2]

[1] Geotechnical and Structural Engineering Research Center, Shandong University, Jinan, Shandong 250061, China.

[2] School of Civil Engineering, Shandong University, Jinan, Shandong 250061, China.

[3] School of Qilu Transportation, Shandong University, Jinan, Shandong 250061, China.

Corresponding author: Bin Liu (liubin0635@163.com)

Key Points:

- Forward modeling for characterizing physical rules was embedded into the network training, and data misfit was used as a loss function.
- Several variants of the network architecture have been designed to fit various data forms under different observation devices.
- Deep learning inversion was successfully applied to field data using a transfer learning method.

## Abstract

The direct-current (DC) resistivity method is a commonly used geophysical technique for surveying adverse geological conditions. Inversion can reconstruct the resistivity model from data, which is an important step in the geophysical survey. However, the inverse problem is a serious ill-posed problem that makes it easy to obtain incorrect inversion results. Deep learning (DL) provides new avenues for solving inverse problems, and has been widely studied. Currently, most DL inversion methods for resistivity are purely data-driven and depend heavily on labels (real resistivity models). However, real resistivity models are difficult to obtain through field surveys. An inversion network may not be effectively trained without labels. In this study, we built an unsupervised learning resistivity inversion scheme based on the physical law of electric field propagation. First, a forward modeling process was embedded into the network training, which converted the predicted model to predicted data and formed a data misfit to the observation data. Unsupervised training independent of the real model was realized using the data misfit as a loss function. Moreover, a dynamic smoothing constraint was imposed on the loss function to alleviate the ill-posed inverse problem. Finally, a transfer learning scheme was applied to adapt the trained network with simulated data to field data. Numerical simulations and field tests showed that the proposed method can accurately locate and depict geological targets.

## Plain Language Summary

The direct current (DC) resistivity survey is a geophysical method for imaging the resistivity model corresponding to an underground medium and solving geological problems. The process of reconstructing the resistivity model from observational data is ill-posed, which can easily lead to incorrect interpretations. In this study, we propose a physics-based deep learning (DL) inversion method that incorporates the law of electric field propagation into a neural network. This method directly approximates the inversion process and builds a nonlinear mapping from the survey data to the resistivity model. The introduction of physical laws can not only alleviate the dependence of DL methods on a large amount of training data but also eliminate the need for real models. Therefore, this method is more suitable for an actual scenario where the real resistivity model is challenging to obtain. This method was verified through numerical simulations and physical model tests. Furthermore, we applied this method to advanced tunnel surveys. In several cases, the inversion results were consistent with the geological conditions present after the excavations.

## 1 Introduction

The direct-current (DC) resistivity method is one of the most commonly used solutions in geophysical surveys (Slater, 2007; Loke et al., 2013). This method is characterized by low economic cost, high survey efficiency, and strong sensitivity to water-bearing structures. It has been widely used for various purposes, including traffic engineering (Donohue S et al., 2011; Guo et al., 2019; Pang et al., 2022), dam surveys (Kim et al., 2007; Bedrosian et al.,2012), and environmental engineering (Coscia et al., 2011; Bellmunt et al., 2016). An effective inversion method is key to improving the reliability of imaging techniques. Currently, linear inversion is the mainstream method used for inverting real-world data. It predicts model per the physical laws of geoelectric fields. However, its inversion results are highly dependent on the initial model, and the local optimal solution is obtained using this method. Its imagining results usually contain artifacts that influence the geological interpretation of the results.

Using nonlinear inversion methods can mitigate such problems because of their strong capacity for global search. Applications of nonlinear inversion methods in inverting DC resistivity data have been widely studied, including the use of genetic algorithms (Liu et al. , 2012), ant colony algorithms (Zhang & Liu, 2011), and simulated annealing algorithms (Santos et al.,2006) to obtain globally optimal solutions. However, these methods have not been popularized for inverting real-world data because of their slowness in outputting solutions. In contrast, Neural network methods operate relatively quickly and enable well-fitted nonlinear mapping between input data (potential or apparent resistivity) and output data (resistivity model) by extracting information from large training sample sets (Singh et al. , 2010; Neyamadpour et al., 2010; Jiang et al., 2018). The method is only time-consuming during training, and the trained networks have an extremely high inversion efficiency during inference, which makes them suitable for resistivity surveys. In recent years, with the significant optimization of artificial intelligence algorithms and computing performance, an upgraded version of neural networks, deep learning (DL), has rapidly developed (Hinton & Salakhutdinov, 2006). DL has a greater ability to construct complex nonlinear mappings than older neural networks methods. Solving geophysical inversion problems using DL has gradually become a research hotspot (Araya-Polo et al., 2018; Wu et al., 2019; Puzyrev, 2019; Huang et al., 2021). Currently, research on DL inversion for real-world data from electrical surveys is still in the exploratory stage. Liu et al. (2020) studied DL inversion using convolutional neural networks (CNNs) for resistivity survey data. In DL inversion methods, depth weighting and smoothing constraints are added to the loss function to alleviate the ill-posed problems. For synthetic data, the neural network achieves better inversion results than traditional linear inversion without involving linearization theory. Based on this, a variable convolution kernel is used to adapt the apparent resistivity image features, further improving the imaging performance of complex models (Liu et al., 2021). Aleardi et al. (2021) combined CNN inversion with a Monte Carlo simulation framework to estimate model uncertainty caused by noisy data. Vu and Jardani (2021) extended DL inversion to 3D surface survey imaging.

Currently, most DL-based resistivity inversions are purely data-driven and trained in a supervised manner, making their performance heavily dependent on an extensive training set. This poses two challenges when applying these methods to real-world data. (1) Resistivity model information corresponding to real geological models is difficult to collect. In other words, supervised training with real-world data and corresponding models is difficult. A commonly used solution is to train networks with a synthetic dataset and fine-tune them with a few field samples to adapt them to real scenarios; this method is called, transfer learning. (2) Even for synthetic datasets, the resistivity model covering generic realistic exploration scenarios should be massive and exhaustive. Owing to limitations in time and computing resources, this type of synthetic dataset is difficult to generate. Deep neural networks trained only by a dataset with insufficient samples may not be able to accurately approximate nonlinear mapping between survey data and resistivity models. In other words, the networks may not comply with the physical laws of inversion (e.g., electric field propagation). Therefore, DL based on physical laws is more promising for the inversion of DC resistivity data. This idea has been adopted for seismic inversion problems such as those outlined by Jin et al. (2021). They embedded a forward modeling module at the end of a neural network to form a data misfit and realize unsupervised learning. Colombo et al. (2021) applied unsupervised learning inversion to transient electromagnetics and obtained high-resolution resistivity models for synthetic and field data. Liu et al. (2022) incorporated the physical laws of magnetotelluric wave propagation into a purely data-driven DL approach and successfully applied this method to field data.

To the best of our knowledge, an unsupervised DL method based on physics has not been reported for DC resistivity data inversion. This novel method, driven by physical laws, is expected to have promising applications in resistivity survey data. However, there are three crucial issues related to using physics-driven unsupervised DL for DC resistivity inversion that must be addressed: (1) How can the laws governing electric field propagation be adopted in deep neural networks to achieve unsupervised learning? (2) How can prior information constraints be used to ensure the convergence of the network training process? (3) How can the method be applied to field data when the size of the real-world training dataset is too small to support the network training?

In this study, to address the first problem listed above, we constructed a physics-driven resistivity data inversion network (PhResNet) that combines the architectures of widely used CNNs and fully connected neural networks (FCNNs). In this network, the forward operator simulating electric field propagation is applied to the inversion architecture, which helps guide the network training by fitting the forward results of the prediction model with the survey data. On this basis, a dynamic smoothness constraint is imposed on the loss function to solve the second problem listed in the previous paragraph. Finally, we proposed the use of a transfer learning method for a small amount of resistivity survey data to solve the third problem listed in the previous paragraph. PhResNet was successfully applied to real-world data collected during an advanced tunnel survey. The inversion results of the field tests matched the excavation disclosures, validating the feasibility and effectiveness of PhResNet.

## 2 Methods

In this section, we present a physics-based network for unsupervised resistivity inversion (PhResNet). First, we designed the overall network architecture to enable unsupervised learning. Subsequently, we designed two specific encoders for different types of DC resistivity data. To obtain an effective inverse network, we proposed a dynamic smoothing constraint to guarantee training convergence. Finally, we improved the transfer learning method by applying unsupervised inversion to real-world data.

### 2.1 Overall Architecture of PhResNet

The objective of this study was to achieve unsupervised learning by introducing physical laws. The goal was to eliminate the dependence on labels (resistivity model) and improve the generalization ability of the inversion network. For DC inversion, the physical law is represented by an electric field distribution obeying Poisson's equation. Forward modeling can transform a geoelectric model into observational data using Poisson's equation and boundary conditions. Therefore, forward modeling was added to the neural network as an effective way to introduce the physical laws of electric field propagation. Among the existing unsupervised learning methods, self-supervised learning using only observed data is common. In this method, forward modeling enables the mapping process from the output model to the data. We used data misfit as the loss function after the prediction model was mapped to the data. The advantage is that the training process using this loss function does not require labels. Based on the above analysis, forward modeling was added to the network. Finally, a large number of samples were used to train the network parameters through the loss function for the network to fully learn the physical laws.

The training sample set included N groups of resistivity models $\mathbf{m}_i^{\text{Label}}$ ($i \in N$) and their corresponding survey data $\mathbf{d}_i^{\text{obs}}$ ($i \in N$). The neural network parameters were represented by $\mathbf{w}$. The purpose of the inversion network is to construct a mapping $F$ of $\mathbf{d}_i^{\text{obs}}$ to $\mathbf{m}_i^{\text{Label}}$ by training $\mathbf{w}$.

$$F(\mathbf{d}_i^{\text{obs}}, \mathbf{w}) \to \mathbf{m}_i. \tag{1}$$

A typical supervised learning pipeline is illustrated in Figure 1. The resistivity model and survey data were used as the input information for the neural network. The loss function was set to the residuals of the prediction model $\mathbf{m}_i^{\text{pre}}$ and the real model $\mathbf{m}_i^{\text{Label}}$. Network gradients were calculated to update the network parameters.

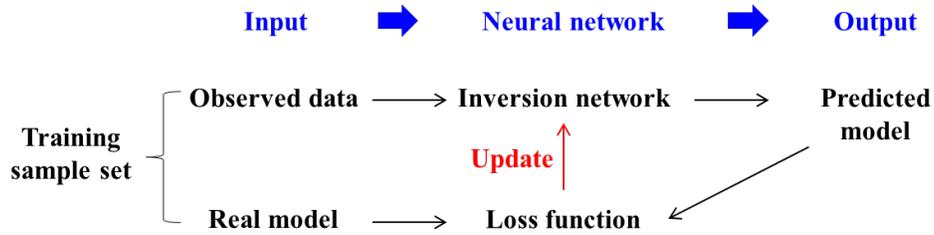

Figure 1. Inversion network framework of supervised learning. The real model and survey data taken in pairs from the training sample set are used as network input. The predicted model is generated from the observed data using an inversion network, and then added to the loss function together with the real model. Finally, the inversion network parameters are updated using the loss function.

However, resistivity models are typically unavailable during field surveys, and supervised inversion networks cannot be trained owing to the lack of resistivity model labels. Therefore, it is difficult for supervised inversion networks to produce good inversion results using real-world data. The inverse network framework for unsupervised learning was designed by adding forward modeling to the inverse network framework (Figure 2). Unlike the inversion network framework of supervised learning, only survey data are available in the training sample set in the case of unsupervised learning. The forward modeling module was placed after the predicted model. The forecast data generated by forward modeling had the same array form as that of the predicted data. The input of the loss function as changed from the model to the data, i.e., the network parameters were updated by fitting the predicted data to the observed data.

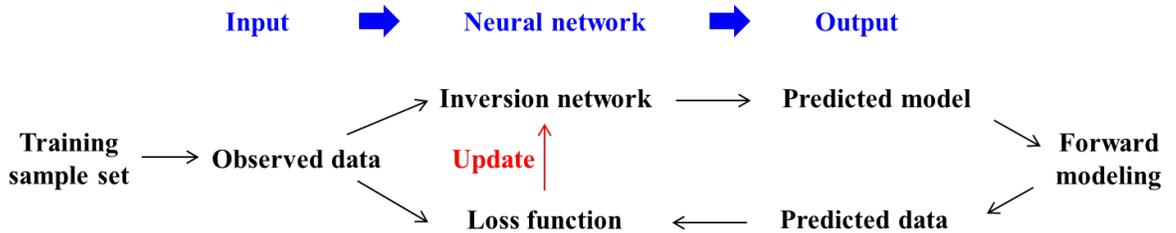

Figure 2. Inversion network framework of unsupervised learning. The survey data taken from the training sample set is used as the network input. The predicted model obtains predicted data using forward modeling. The predicted data and the observed data are

**added to the loss function together. Finally, the inversion network parameters are updated using a data-driven loss function.**

Forward modeling is based on the physical laws of electric field propagation. The governing equation for the propagation of the electric field is shown below (Dey & Morrison, 1979):

$$-\nabla \left( \frac{\nabla \Phi(x, y, z)}{\rho(x, y, z)} \right) = \left( \frac{I}{\Delta V} \right) \delta(x - x_0) \delta(y - y_0) \delta(z - z_0). \tag{2}$$

Equation (2) is a Poisson equation derived from Ohm's law and the conservation of current, which governs the relationship between electrical resistivity $\rho$ and potential $\Phi$. I is the current in amperes and $(x_0, y_0, z_0)$ are the coordinates of the point source of the injected charge. $\Delta V$ is the unit volume and $\delta(\cdot)$ represents the Dirac delta function.

The potential is solved using a resistivity model $\rho$ using governing equations and boundary conditions. This process is called forward modeling in DC resistivity surveys. We applied the finite element method to solve the forward model discretely. This is because the finite element method is more accurate than the finite difference method in dealing with nonuniform continuum and complex shapes. Thus, this method is suitable for target detection.

2.2 Extracting Information from Potential Data

In this section, data feature extraction methods are proposed to adapt to various observation devices in DC resistivity surveys; additionally, the data from different observation devices have different array forms. Data can be roughly classified into image and nonimage data. For example, the apparent resistivity of surface surveys is image data, and the potential surveyed in holes is non-image data.

**PhResNet-i for Image Data**

Image data have a spatial correspondence with the resistivity model (Liu et al., 2020), which is suitable for feature extraction using CNNs. U-Net architecture based on CNNs has good localization and feature representation capabilities (Ronneberger et al., 2015). Therefore, we used U-Net to extract features from the image data. U-Net is usually composed of two parts: an encoder and decoder. In particular, shallow features (encoder part) and deep features (decoder part) were jointly used for inversion using the shortcut. The physics-driven resistivity inversion network of the PhResNet-i image data (PhResNet-i) is shown in Figure 3. First, the network parameters were initialized randomly. Subsequently, multiple apparent resistivity image data were simultaneously fed into the encoder network to extract features through batch processing. A prediction model corresponding to the input data was generated by the decoder network. Furthermore, the prediction data were computed using a forward modeling. Finally, the average gradients of the multiple models were computed using the loss function. This model gradient was back propagated along the red line to update all network parameters.

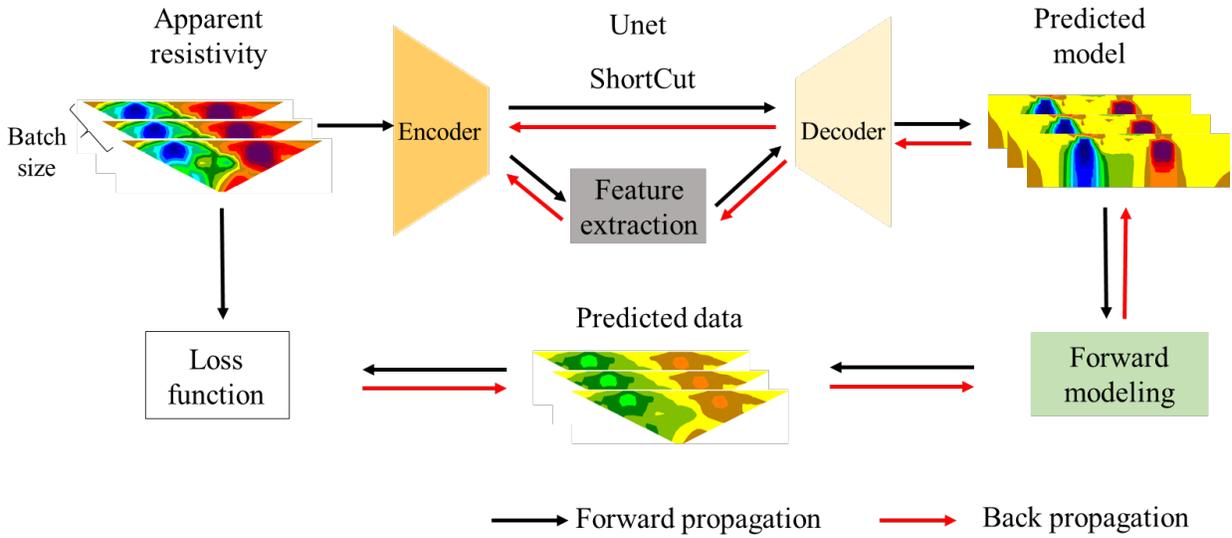

Figure 3. Physics-driven resistivity inversion network for image data (PhResNet-i).

**PhResNet-n for Non-image Data**

It is difficult to construct a spatial correspondence with the resistivity model for non-image data; therefore, CNNs are unsuitable for nonimage data. An FCNN can be used to extract features from non-image data, but its fully connected neurons will result in a large number of parameters and affect the training efficiency. Drawing on the method of splitting data by shot points in seismic DL inversion (Li et al., 2020), we attempted splitting the non-image data (potential) according to current electrodes. Each split dataset was processed using the same network parameters to reduce the total network parameters. However, unlike seismic survey data (time series), each DC resistivity survey data point usually corresponds to four electrodes (two current electrodes and two potential electrodes). That is, each datum contains the information of four electrodes with thousands of location data. This is difficult for the network to learn. As shown in Figure 4, we attempted to address this problem by implementing the following measures:

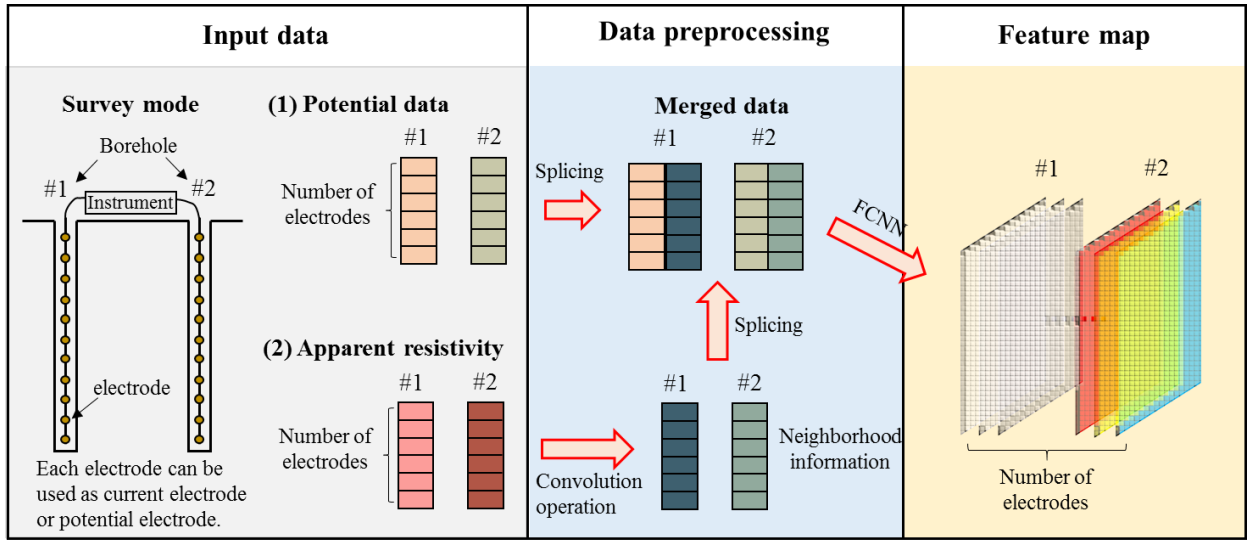

**Figure 4.** Schematic diagram of encoder network operation.

(1) The location information was supplemented by the apparent resistivity data. As shown in the left column of Figure 4, the input data was a collection of resistivity potential and apparent data. The apparent resistivity data incorporating the geometric factor carried spatial information, which was beneficial for the neural network to reduce the search range of the solution. Note that both apparent resistivity and potential data were used as inputs to the neural network, unlike traditional methods that use only one type of input data. Because the values of the apparent resistivity and potential are significantly different, two network input paths were constructed.

(2) Inversion networks distinguish data powered by different holes by grouping them. As shown in the left column of Figure 4, the data for the two holes is divided into two parts. The data split by the current electrode needed to be split twice according to the position of the borehole where the current electrode was located. This is because the values and trends of the data are similar when the current electrode is located in the same borehole. For the groups with current electrodes that were not in the same borehole, the data values and trends were considerably different. Therefore, the data from different boreholes were not processed using the same network parameters.

(3) The information lost after grouping was compensated for with the extracted neighborhood information. The features between different sets of data were ignored after the split operation was performed. To solve this problem, the split data were sorted according to the current electrode's spatial order. The difference in electric fields between adjacent groups was generated by moving the current electrode. This difference in electric field difference was regarded as neighborhood information. As shown below the middle column of Figure 4, we extracted this neighborhood information using convolution operations to supplement more effective information at the network input.

By combining the three measures listed above, an encoder network for non-image data was designed, as shown in Figure 4. Based on cross-hole electrical resistance tomography, the two boreholes were numbered #1 and #2, respectively. Electrodes were placed in the boreholes and used either as current or potential electrodes. Neighborhood information was extracted by the convolution of the apparent resistivity data. Further, the merged data were obtained by splicing the neighborhood information and potential data. Finally, a feature map proportional to the size of the resistivity model was generated after incorporating the fused data into the FCNN. Feature maps contain both low- and high-dimensional data features, based on which the prediction model was directly obtained by inputting the feature map into CNNs.

The physically driven resistivity inversion network for nonimage data (PhResNet-n) that we developed is shown in Figure 5. The main difference between PhResNet-n and PhResNet-i is the use of an FCNN-based encoder network instead of a U-Net. The two network architectures implemented the inversion of image/non-image data. This means that the DL methods were no longer limited to observation devices.

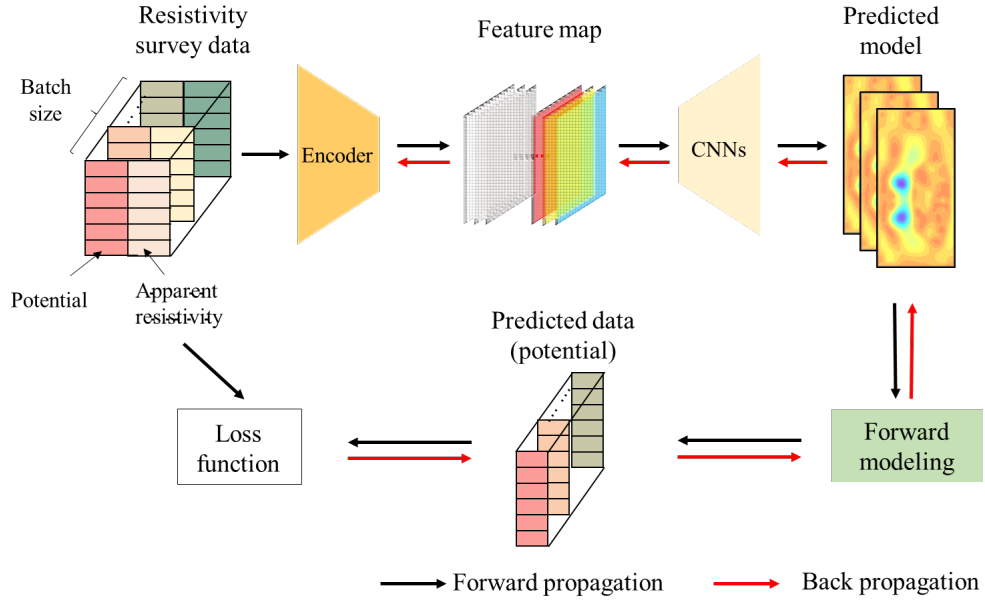

**Figure 5. Physics-driven resistivity inversion network for non-image data (PhResNet-n).**

2.3 Gradient Calculation Based on Dynamic Smooth Constraint

Drawing from traditional linear inversion, a smoothness constraint was added to the loss function to ensure that the network training process converged. The loss function of PhResNet includes a data term and a model term, as follows:

$$Loss = \left(f(\mathbf{m}) - \mathbf{d}^{obs}\right)^T \left(f(\mathbf{m}) - \mathbf{d}^{obs}\right) + \lambda \left(\partial^n \mathbf{m}\right)^T \left(\partial^n \mathbf{m}\right), n = 1 \text{ or } 2 \; . \quad (3)$$

$f(\cdot)$ represents forward modeling mapping. $\mathbf{m}$ is predicted model. $\lambda$ is a regularization factor that balances the data and model terms. The model gradients were solved using the Gauss-Newton method for the above equations.

$$\delta \mathbf{m} = \left(\mathbf{J}^T \mathbf{J} + \lambda \mathbf{C}^T \mathbf{C}\right)^{-1} \mathbf{J}^T \left(f(\mathbf{m}) - \mathbf{d}^{obs}\right), \quad (4)$$

where $\mathbf{J}$ is the Jacobian matrix and $\mathbf{C}$ is the smooth constraint matrix.

Smooth constraints are double-edged swords. Weak constraints may not guarantee convergence in the inversion process. Although strong constraints may alleviate the multi-solution problem of inversion, its predicted model was too smooth and could not accurately reflect the abnormal areas. To solve this problem, we attempted to use a gradient-calculation strategy with dynamic-smoothness constraints. In the early stage of network training, the gradient calculation process was unstable because the predicted model was noticeably different from the real model. Therefore, the smoothness constraint was enhanced by a larger regularization factor, $\lambda$. Furthermore, the influence of smooth constraints was reduced in late training to achieve accurate imaging of the target regions. The dynamic regularization factor calculation formula is as follows:

$$\lambda = \lambda_0 \times (1.0 - \text{epoch}/\text{max\_epoch})^{\mu}. \tag{5}$$

Where $\lambda_0$ is the initial value, $\text{max\_epoch}$ is the maximum number of training times, and $\mu$ is the rate of change factor.

2.4 Transfer Learning

In a real resistivity survey, the number of DC resistivity survey data cannot meet the requirements of network training. Transfer learning is a common method in DL to solve this issue. We tried to devise a new transfer learning method suitable for a small amount of DC resistivity survey data. The network parameters trained by the synthetic samples were used as the starting point of the transfer learning process, followed by fine-tuning a part of the network parameters with a small number of real samples.

There are two popular methods for transfer learning: (1) full fine-tuning, i.e., updating the parameters for all layers of the network. (2) Linear probing: retraining the last linear layer. Full fine-turning can improve the feature extraction of pre-trained networks using real-world data, whereas linear probing directly inherits the feature extraction method of synthetic data, and may not be effectively applied to real-world data. Full fine-tuning generally has higher accuracy than linear probing (Kornblith et al., 2019; He et al., 2020). However, for the problem of DC resistivity survey data, the search range of the pre-trained network relies on training samples with synthetic data. If the deviation between the real resistivity model and trained samples is large, the inversion results cannot easily approach the real model. Linear probing has a better global search ability than full fine-tuning, which was expected to alleviate the above problems. Therefore, we devised a transfer learning strategy that combined the advantages of full fine-tuning and linear probing, a schematic of which is shown in Figure 6. First, the pre-trained network was linearly explored using real-world data to expand the search for solutions. Second, the network was fully fine-tuned to improve feature extraction from real-world data. Finally, the network was iteratively trained and inverted on the new input (target data) multiple times until a set number of times or a convergence value was reached. The proposed transfer learning strategy had two positive outcomes: 1) the network parameters were further adjusted to better adapt to the target data; 2) the optimized network was enabled to generate more accurate resistivity models.

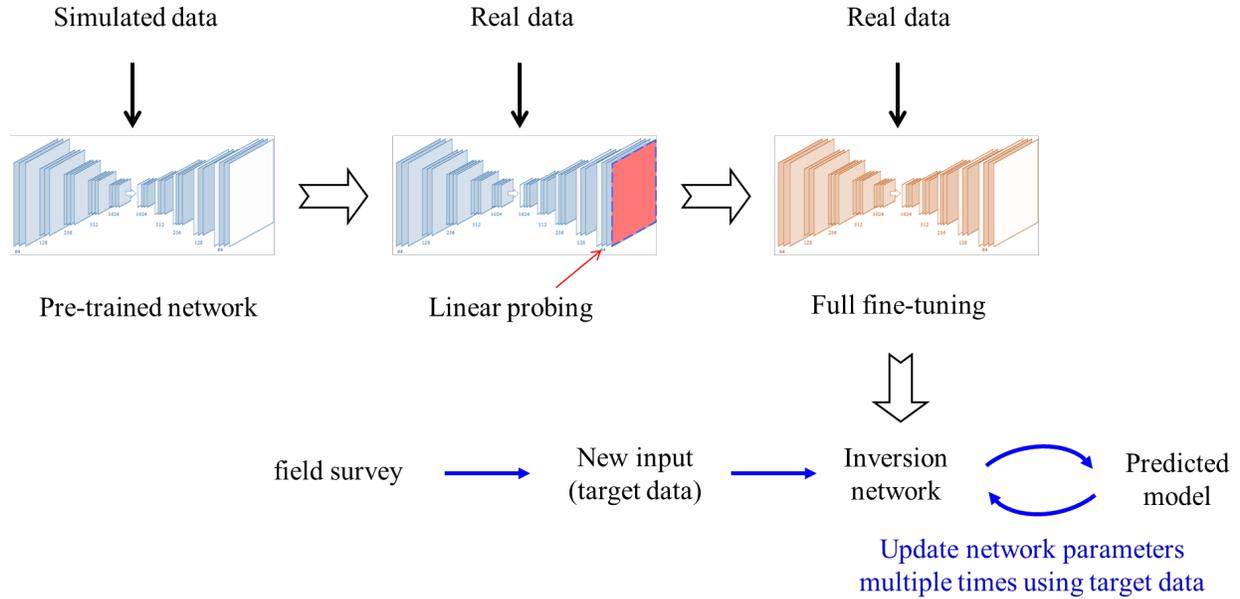

**Figure 6. The transfer learning flow chart for the real-world data of electrical method**

## 3 Synthetic Inversion Tests

In this section, we compare the proposed PhResNet with traditional linear inversion methods using numerical tests. We did not make comparisons with other deep learning methods because they all require supervised training using real resistivity models.

### 3.1 Training details

We employed the SGD (Ruder S, 2016) optimizer with a momentum parameter $\beta_1=0.9$ and a weight decay of $1\times10^{-4}$ to update all parameters of the network. The initial learning rate was set to $0.2\times10^{-4}$. The size of the minibatch was set to 8. All network layers were initialized randomly. During the network training, dropout techniques were used to avoid overfitting the training data. The implementation of network training was conducted using Pytorch on a desktop system (Intel(R) Xeon(R) Gold6148 CPU @2.40GHz, 512GB RAM. GPU: NVIDIA TITAN RTX). This configuration was used for all the tests conducted in this study.

### 3.2 3D Surface ERT

A synthetic dataset for a 3D surface survey was built using random disturbance and discrete combinations. The inversion area consisted of a grid of 2 m×2 m×2 m, with a range of 16 m(X)×66 m(Y)×16.5 m(Z). Four survey lines were placed at a distance of 4 m from each other. In total, 128 electrode points, each separated by 2 m, were used. The background resistivity of the resistivity model was 1000 Ohm·m. Figure 7 shows the possible shapes of the low-resistance target from a left-hand side view. Its resistivity was 20 Ohm·m. To obtain sufficient information, we used observation devices including the Wenner-Schlumberger, Dipole-Dipole, and Pole-Dipole. The dataset contained a total of 4626 samples. It was divided into training, validation, and test sets in a ratio of 10:1:1.

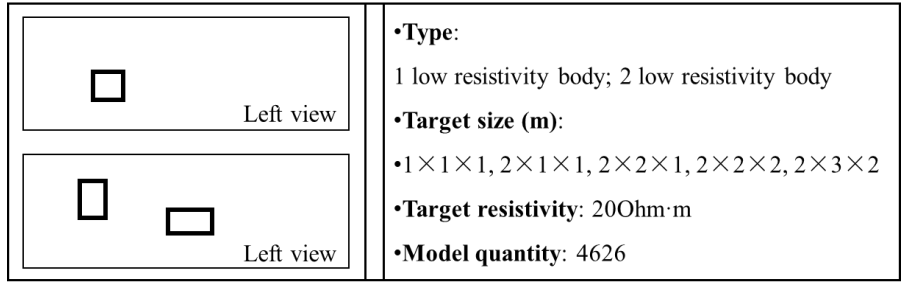

**Figure 7. Schematic and parameters of models for 3D surface survey.**

The inversion results are shown in Figure 8. It comprised three parts: the real resistivity model, the prediction model, and slice map. The predictive model obtained using the linear approach showed only one target. This may be because the signal of the shallow target was stronger than that of the deep target, causing the deep signal to be masked. PhResNet-i has strong information mining ability and can extract information effectively from weaker signals. Consequently, both targets could be described and differentiated. The resistivity value of the target area was close to the actual value.

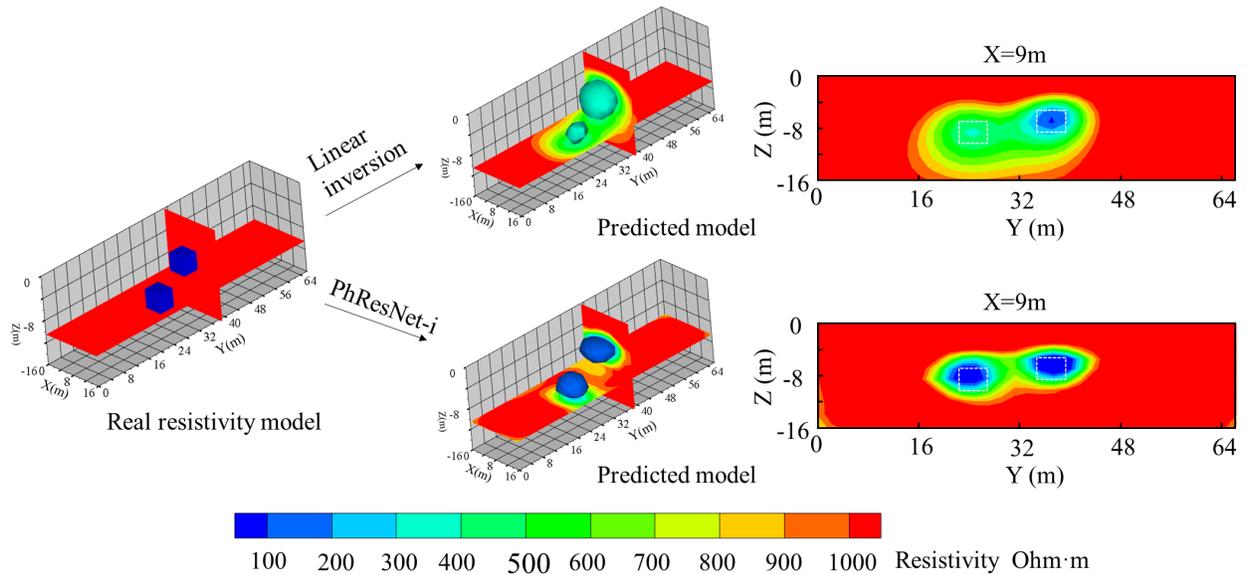

**Figure 8. Inversion comparison test of geoelectric model with two targets.**

### 3.3 2D Cross-hole ERT

Similar to that described in the previous section, a synthetic dataset for a 2D cross-hole ERT was built. The inversion area consisted of a grid of 1 m×1 m, with a range of 16 m(X)×32 m(Z). Two survey lines were 14 m apart from each other. A total of 64 electrode points with a spacing of 1 m were used. The background resistivity of the resistivity model was 200 Ohm·m. Figure 9 shows the possible shapes of the low-resistance target in a left-hand view. Its resistivity was 20 Ohm·m. To obtain sufficient information, we used the following observation devices: a bipole-bipole, a dipole-dipole, and a pole-pole. The potential and apparent resistivity data were obtained using forward modeling. The dataset contained 4880 samples. It was divided into training, validation, and test sets according in the ratio 10:1:1.

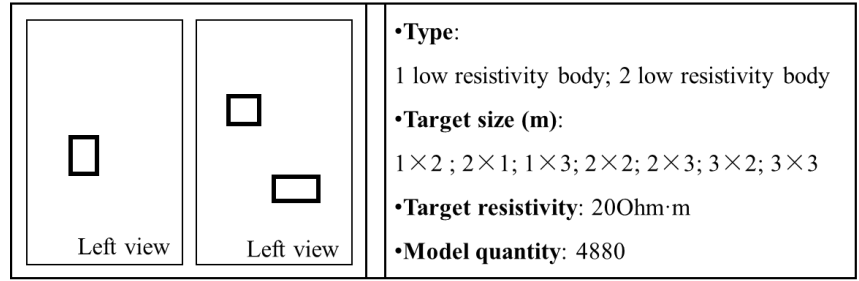

**Figure 9. Schematic and parameters of models for 2D Cross-hole survey.**

The cross-hole survey data were processed using PhResNet-n because they contained non-image data. Figure 10 shows the inversion results for two low-resistivity targets that were close to each other. Horizontal resolution using cross-hole ERT is generally poor. When the horizontal distance between two targets at the same depth is small, the imaging results of the two targets can be easily combined using linear inversion. Because the inversion process falls into a local optimum, it is difficult to jump out of the local optimum, even if the number of iterations is increased. The unsupervised inversion network PhResNet-n had a strong global convergence ability because it was trained using a large number of samples. PhResNet-n can accurately discriminate between and image multiple objects. As shown in Figure 10, the results obtained using PhResNet-n were close to the actual model in terms of size, shape, and resistivity values.

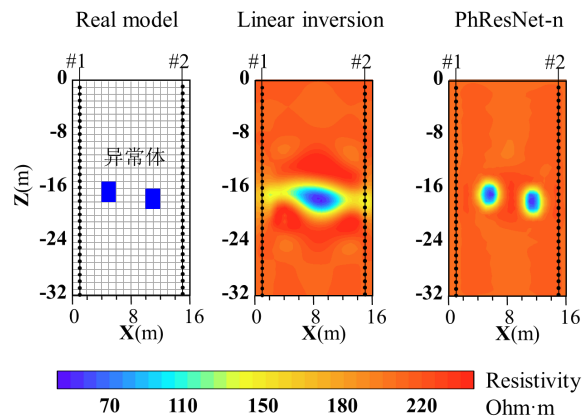

**Figure 10. Inversion comparison test using cross-hole ERT with two closely targets.**

## 4 Model Test

Unsupervised learning inversion was applied to real-world data using the transfer learning method described in Section 2.4. The specific process followed was: The modeling parameters were determined according to the electrode coordinates, observation devices, and detection requirements. Training datasets with large amounts of synthetic data were constructed based on the geology and the potential anomalies of the surveyed area. ResinvNet was pre-trained on synthetic datasets to obtain the inversion capabilities for synthetic data. Based on this, the network was retrained using the actual data from the previous stage. Finally, the trained PhResNet was applied to the new data. The final resistivity model was generated after multiple iterations.

4.1 Data Preparation

We designed a model test to invert the low-resistivity anomalies using cross-hole ERT. The low-resistivity body is designed to be the simplest block rather than a complex shape. The application of unsupervised learning inversion to real-world data is still at the exploratory stage. The number of abnormal bodies found was 1 or 2. The ratio of the model to the engineering prototype was 1:20. A comparison of geometric factors is presented in Table 1.

Table. 1 Comparison of different geometric factors of the model in prototype and experiment.

| Geometric factors | Engineering prototype (m) | test model (m) |
|---|---|---|
| Model size | 16×16×32 | 0.8×0.8×1.6 |
| Borehole depth | 32 | 1.6 |
| Borehole spacing | 16 | 0.8 |
| Electrode spacing | 1.0 | 0.1 |

The model design is shown in Figure 11(a) and the corresponding photographs are shown in Figure 11(b). The survey area was 0.8 m (length) × 0.8 m (width) × 1.6 m (depth). It was connected to the ground on all four sides, which helped mitigate the boundary effects in the electrical tests. Four survey lines were arranged in the survey area. A total of 4×16 electrodes were used, with a spacing of 0.1 m on the same survey line. For better coupling, the electrodes were packed with wet clay. The survey area was filled with fine-grained soil. The background resistivity ranged from 200~400 Ohm·m.

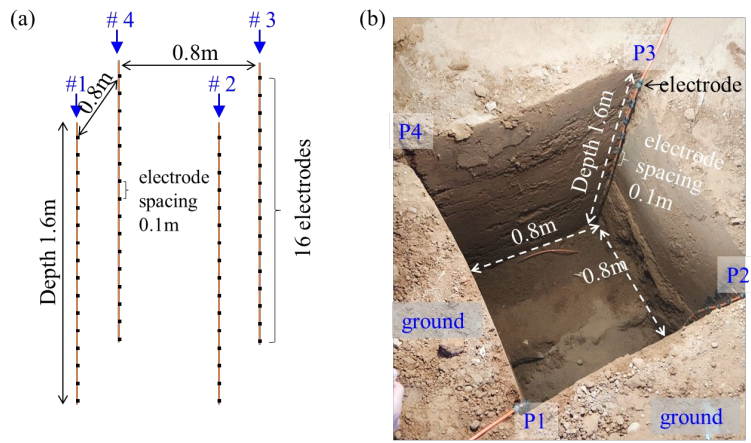

Figure 11. Schematic diagram of model test of aquifer detection

Clay, salt, and water were used to simulate anomalous bodies. The size of each abnormal body was 0.2 m× 0.1 m× 0.1 m. Its resistivity value was maintained at 20~50 Ohm·m by controlling the material ratio. The final imaging area consisted of the two diagonal faces of the survey area (#1–#3 and #2–#4). Fifty sets of survey data were collected by placing the anomalous bodies horizontally or vertically at different depths. A photograph of the test process is presented in Fig. 12.

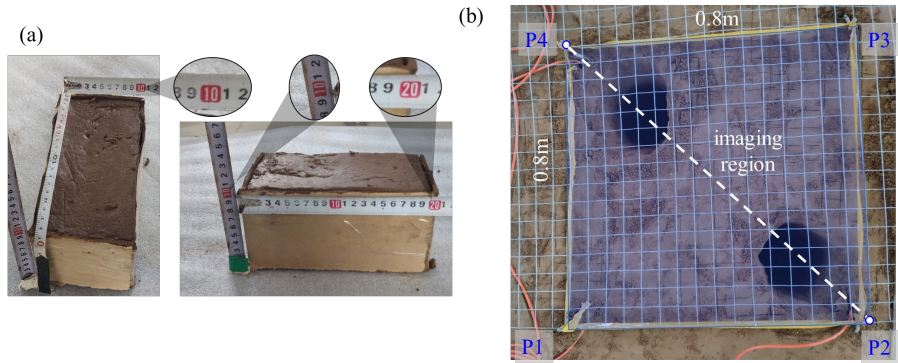

**Figure 12. Photos of abnormal body filling process.**

4.2 Results

The imaging results obtained using unsupervised learning inversion and linear inversion are shown in Figure 13. The black lines on both sides are the survey lines (located at X=0 m and X=1.15 m). The black dots on the measurement line represent the electrodes. The white dotted box indicates the actual location of the anomalous body . The results of the linear inversion could not differentiate the two anomalies. PhResNet-n effectively located and imaged the two abnormal bodies with an error of only 0.1 m. In contrast, the imaging results obtained using PhResNet-n were relatively close to real model.

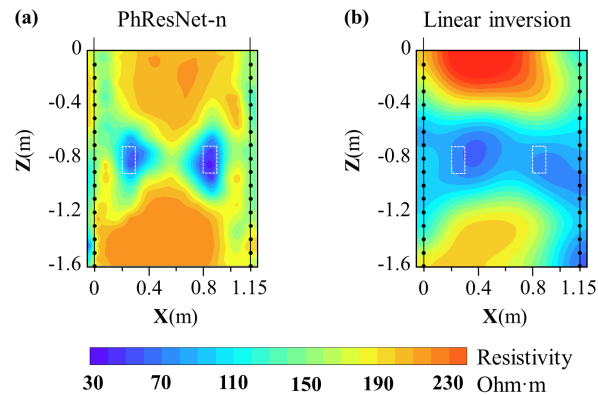

**Figure 13. Comparison of inversion results using different models. (a) Unsupervised learning inversion (PhResNet-n), (b) Linear inversion.**

## 5 Field survey

We conducted a field-test application study to verify the effectiveness of the unsupervised learning inversion method in practical engineering.

5.1 Engineering overview and detection scheme

The survey area was located in a water diversion project in northwest China. The tunnel was excavated using a drill-and-blast method. The survey site in the tunnel was located at the bottom of the river. The top of the tunnel was approximately 271 m from the riverbed. There are many faults in this area, and groundwater is recharged by rivers. Therefore, the area is prone to water inrush disasters during tunnel excavations. Located 12 m in front of the survey site, water flow occurs in an advanced borehole. The water flow rate reached 1300 m³/h. This water may have originated from fissure water because the surrounding rock was intact. The water gushing speed decreased after full curtain grouting. After the slurry solidified, the tunnel continued to be excavated to the survey site. We performed cross-hole ERT using probe holes to identify potential adverse geology.

A schematic of the advanced survey is shown in Figure 14. The tunnel face was 7.8 m×7.8 m. Probe holes were designed in the corners to increase the imaging range. The distances in the horizontal and vertical directions of the probe holes were 5.0 m and 4.0 m, respectively. The probe holes were named H1-H4 clockwise. A total of 16×4 electrodes were used, with a spacing of 1.0 m. The survey lines penetrated the probe hole using a pvc tube.

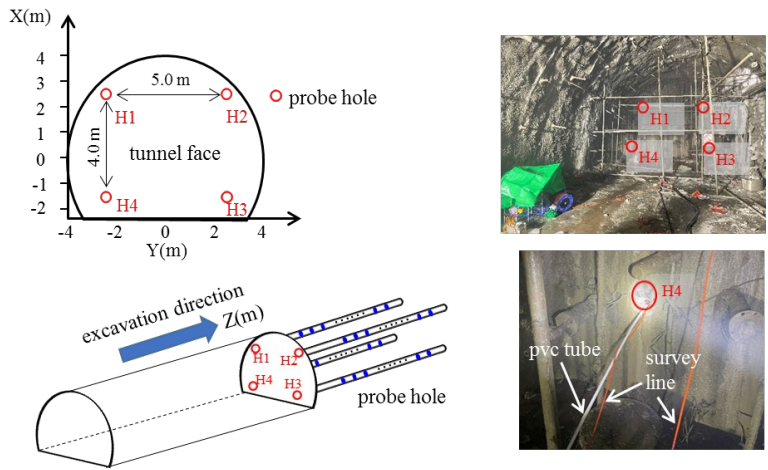

**Figure 14. Schematic diagram of advanced survey and field scene.**

5.2 Inversion and excavation results

The noise level of the data was high owing to the complex tunnel environment. Therefore, the regularization factor of the smooth constraint was maintained at a high value to guarantee convergence. The inversion results are shown in Figure 15. In the range of 0~8 m, the resistivity value of the geological body was high but lower than the normal resistivity value of dry rock. The corresponding excavation site is shown in Figure 16(a). The surrounding rock in this area was wet. We speculate that the area was filled with water in the early stage, but the hidden danger of the water effluent was eliminated after grouting treatment. In the range of 8–10 m, there are distinct low-resistivity regions in both the imaging results (H1-H3 and H2-H4). A small fault was discovered at this location during an earlier investigation. The fault was then treated using approximately 150 t of grouting. Therefore, we speculate that fissure water still exists at this location. The corresponding excavation site is shown in Figure 16(b). We observed consolidated slurry and fissure water, which is consistent with the inversion results. There was no obvious low-resistivity anomaly at 10~16 m. The area was excavated later, with no water coming out.

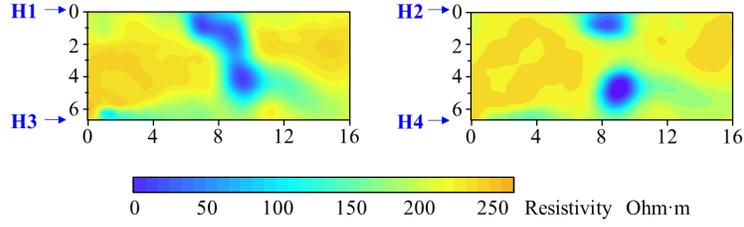

**Figure 15. The inversion results of advanced cross-hole ERT.**

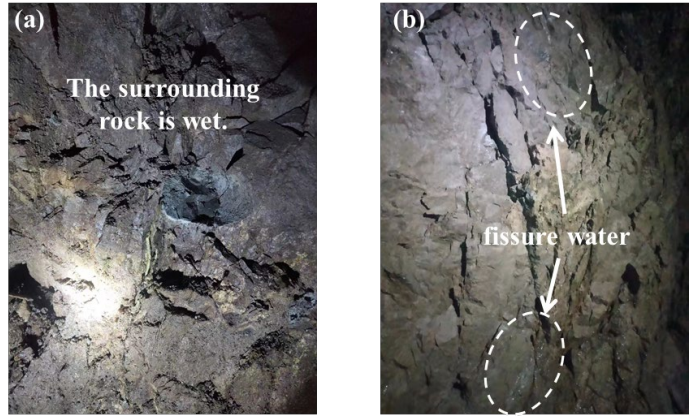

**Figure 16. Geological conditions revealed by excavation**

It should be noted that the effect of transfer learning is limited because of the lack of data in the early stages of the project. This rendered the unsupervised inversion network insufficient for processing real-world data. In addition, owing to the high level of data noise, the imaging resolution was low. The inversion result could only roughly reflect the range of the abnormal body but could not be imaged in detail.

## 6 Discussion

Unlike the model misfit in the original method, the data misfit is more beneficial for the network to learn the physical laws. As shown in **Equation 6**, the partial derivatives of the loss function based on the model misfit to the model parameters are independent of each other: In other words, the update of each model parameter is free. Training is prone to overfitting owing to the lack of constraints.

$$\frac{\partial L\_m}{\partial \mathbf{m}_i} = \frac{\partial \left( \left( \mathbf{m}_i - \mathbf{m}_i^{\text{Label}} \right)^T \left( \mathbf{m}_i - \mathbf{m}_i^{\text{Label}} \right) \right)}{\partial \mathbf{m}_i} \quad (6)$$

As shown in **Equation 7**, $N_{data}$ represents the total amount of data: the partial derivatives of the model parameters based on the loss function of the data misfit are related to all the data. The data distribution was constrained by the physical laws of the electric field. In other words, the updates of the model parameters were indirectly constrained by the physical laws of the electric field. Therefore, the network is expected to learn the physical laws of the electric field driven by a data misfit through a large number of training samples.

$$\frac{\partial \text{Loss}}{\partial \mathbf{m}_i} = \sum_{k=1}^{N_{data}} \frac{\partial \text{Loss}}{\partial \mathbf{d}_k} \frac{\partial \mathbf{d}_k}{\partial \mathbf{m}_i} = \sum_{k=1}^{N_{data}} \frac{\partial \text{Loss}}{\partial f(\mathbf{m})_k} \frac{\partial f(\mathbf{m})_k}{\partial \mathbf{m}_i} \qquad (7)$$

## 7 Conclusions

In this study, we developed a new inversion method based on unsupervised learning to process the DC resistivity survey data. This method uses a data misfit as a loss function to guide the training of the inversion network by embedding physical rules (of forward modeling) into the network structure. In addition, a dynamic smoothness constraint was added to the loss function to stabilize the training process. Based on this, a transfer learning method was proposed to improve the ability of the inversion network to deal with complex realistic exploration scenarios. The results of the numerical simulations and model tests demonstrate that unsupervised learning inversion can accurately reconstruct nonlinear mapping from the input (potential or apparent resistivity) to the output (resistivity model).

Compared with the existing DL-based resistivity inversion methods, the method we devised eliminates the dependence on the real resistivity model in the training set. Considering that a real resistivity model is difficult to obtain, this method is more suitable for real survey scenarios. Compared to the traditional linear method, the proposed method has two advantages: (1) The inversion time of the trained network is only 1 s. An efficient processing speed can satisfy engineering requirements, especially for large-scale surveys. (2) It has a global search ability and no longer relies on the initial model. However, with a limited number of studies, it is too early to conclude that unsupervised learning inversion is superior to traditional linear methods. Furthermore, network training requires millions or even billions of forward models, whereas the linear method requires only dozens of forward models. Such an efficiency disadvantage can be mitigated using computer hardware and algorithms.

Real survey data need to be accumulated for future studies. The effect of transfer learning was improved by building a high-quality sample library. Finally, we believe that this method can be effectively applied to real survey scenarios.

**Acknowledgments**

This study is funded by the National Natural Science Foundation of China (No. 52021005), the Outstanding Youth Foundation of Shandong Province (No.ZR2021JQ22), the Taishan Scholars Program of Shandong Province of China (tsqn201909003), the Shandong Provincial Natural Science Foundation (ZR2022MF265), and the Young Scholars Future Plan of Shandong University.